\documentclass[a4paper,11pt]{iopart}

\usepackage[utf8]{inputenc}
\usepackage{amsfonts}
\usepackage{amssymb}
\usepackage{amsthm}
\usepackage{cite}
\usepackage[T1]{fontenc}
\usepackage[dvips]{graphicx}
\def\hs{h_{\scriptstyle s}}
\def\hr{h_{\scriptstyle r}}
\begin{document}

\title[Stochatic differential equations with time-delayed feedback and multiplicative noise]
      {Stochatic differential equations with time-delayed feedback and multiplicative noise}
\author{S\'{\i}lvio R.~Dahmen$^{1,2}$ and Haye Hinrichsen$^1$}

\address{$^1$ Universit\"at W\"urzburg, Fakult\"at f\"ur Physik und Astronomie,\\
	 Am Hubland, D-97074 W\"urzburg, Germany\\}
\address{$^2$ Instituto de F\'{\i}sica da UFRGS\\
	Av. Bento Gon{\c c}alves 9500, 91501-970 Porto Alegre RS, Brazil}

\begin{abstract}
\noindent
The stochastic differential equation $\dot{x}(t) = ax(t) + bx(t-\tau) + c\,x(t)\,\xi(t)$ with a time-delayed feedback and a multiplicative Gaussian noise is shown to be related to Kardar-Parisi-Zhang universality class of growing surfaces.
\end{abstract}

%\submitto{J. Phys. A}
% Explanation of PACS numbers:
% 05.50.+q: Lattice theory and statistics
% 05.70.Ln: Nonequilibrium and irreversible processes
% 64.60.Ht: dynamic critical phenomena                                

\parskip 2mm
\vglue 5mm
%============================================================================================
\section{Introduction}
%============================================================================================

Differential equations with a time-delayed feedback have attracted a lot of attention in various fields in recent years~\cite{strogatz,wuensche,cho,boccaletti,pikovsky,argyris,bratsun}. Due to their non-Markovian dynamics such equations exhibit very interesting phenomena. However, our understanding of their mathematical properties is still at the beginning~\cite{hale,amann}. This applies not only to deterministic systems which are more frequently studied but also to \textit{stochastic} differential equations with a time-delayed feedback.

Most of the differential equations with a time-delayed feedback that have been discussed so far may be written in the form
\begin{equation}
\dot{x}(t) = F[x(t)] + G[x(t-\tau)]\,,
\end{equation}
where $F$ and $G$ are certain functions. Here the quantity $x(t)$ may be understood as a one-dimensional stream of data that evolves according to a non-local dynamics. However, slicing the time line into equidistant intervals of width $\tau$ and arranging these segments line by line in a two-dimensional plane, the time-delayed feedback becomes local as it connects points of two neighboring intervals at the same relative position. Combined with an appropriate continuum limit this means that a one-dimensional system with a time-delayed dynamics can be mapped onto a 1+1-dimensional system on a strip with local dynamics. This correspondence works also for stochastic systems, where $F$ or $G$ involve random variables.

%%%%%%%%%%%%%%%%%%%%%%%%%%%%%%%%%%%%%%%%%%%%%
\begin{figure}[t]
\centering\includegraphics[width=120mm]{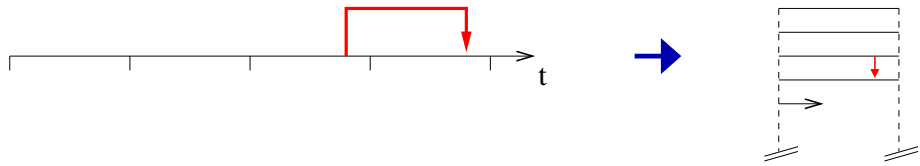}
\caption{\label{fig:mapping}\small
Mapping a one-dimensional time series with a constant time-delayed feedback onto a 1+1-dimensional strip. The time line is sliced into equal intervals of width $\tau$ which are arranged line by line on a strip with spiral-like boundary conditions. As can be seen, the non-local feedback (as examplified by the red arrow) turns into a local interaction between neighboring segments.
}
\end{figure}
%%%%%%%%%%%%%%%%%%%%%%%%%%%%%%%%%%%%%%%%%%%%%

The purpose of this work is to show that this mapping can be used to relate certain one-dimensional time-delayed stochastic differential equations to well-known 1+1-dimensional stochastic models with local dynamics and thereby predict their statistical properties. As an example we study the stochastic differential equation
\begin{equation}
\label{eq:sdeq}
\dot{x}(t) = a \,x(t) + b \,x(t-\tau) + c\,x(t)\,\xi(t)\,,
\end{equation}
where $x(t)\in\mathbb{R}$ is the one-dimensional time series, $a,b$ and $c$ are real-valued parameters, $\tau>0$ is the temporal delay
and $\xi(t)$ denotes a Gaussian white noise with the correlation
\begin{equation}
\label{eq:noise}
\langle \xi(t) \xi(t') \rangle = \delta(t-t')\,.
\end{equation}
Recently this equation was studied numerically by Brettschneider~\cite{Brett}, who observed extremely strong fluctuations of $x(t)$. By relating this system to a 1+1-dimensional stochastic differential equation we will show that these fluctuations can be described in terms of the Kardar-Parisi-Zhang (KPZ) universality class~\cite{KPZ}.

Since $\xi(t)$ is multiplied by the field $x(t)$, the noise in Eq.~(\ref{eq:sdeq}) is said to be multiplicative. As in any system with multiplicative noise the iteration scheme has to be specified. Throughout this paper we shall assume that Eq.~(\ref{eq:sdeq}) is iterated in the It\^o sense. 

Before starting we note that the stochastic differential equation~(\ref{eq:sdeq}) has the following symmetry properties:
\begin{itemize}
\item[(a)] The equation is linear and invariant under translations in time.
\item[(b)] A change of the time scale $t \to \lambda t$ rescales the parameters by
\begin{equation}
a \to \lambda a\,,\quad
b \to \lambda b\,,\quad
c \to \sqrt{\lambda}  \,c\,,\quad
\tau \to \lambda \tau
\end{equation}
\item[(c)] Since the noise is statistically symmetric under reflections, the equation is invariant under a change of sign $c \to -c$. 

\item[(d)] Under the transformation $x(t) \to  e^{\mu t} x(t)$ the parameters change as
\begin{equation}
a \to a-\mu\,, \quad
b \to b \, e^{-\mu \tau}
\end{equation}
\end{itemize}
In what follows we use the invariances (b) and (c) to set $c=1$ without loss of generality. Moreover, the property (d) can be used to fix the parameter~$a$. However, for the sake of clarity we keep $a$ as a free parameter throughout the calculations.

Since the stochastic differential equation invokes a temporally delayed feedback $\tau$, the initial state of the evolution is specified by the whole history of $x(t)$ in a finite temporal interval of width $\tau$, for example $t \in [0,\tau)$. For simplicity we shall assume that $x(t)=1$ on this interval. The question posed here is how $x(t)$ is distributed for large $t$.

Unless stated otherwise we assume that the parameter $b$ is non-negative. This ensures that $x(t)$ with a positive initial condition cannot change sign. To see this consider the transformation $x(t)=e^{y(t)}$ which leads to the stochastic differential equation $\dot{y}(t)=a + b\,e^{y(t-\tau)-y(t)}+\xi(t)$ with a non-multiplicative noise and the initial condition $y(t)=0$ for $0 \leq t < \tau$. Obviously, for $b\geq 0$ this differential equation cannot evolve towards $y(t)=-\infty$ within finite time, hence $x(t)$ is positive for all $t$.

The paper is organized as follows. In Sect. 2 the stochastic differential equation~(\ref{eq:sdeq}) is discretized and integrated numerically. In Sect. 3 we show how the one-dimensional iteration with non-local delay can be mapped onto a two-dimensional system with a local update rule. Carrying out the continuum limit this update rule turns out to be equivalent to a KPZ equation. Finally in
Sect. 4 we compare the theoretical prediction with the numerical results. We finish the paper with some conclusions.

%============================================================================================
\section{Numerical analysis}
%============================================================================================
%
To study Eq.~(\ref{eq:sdeq}) numerically we discretize the time variable $t=nh$ in steps of $h$. Then (using the It\^o scheme) the discretized version of Eq.~(\ref{eq:sdeq}) reads
\begin{equation}
\label{eq:original}
x_{n} \;=\; (1+ah)x_{n-1} + bh\,x_{n-k}+\sqrt{h}\, x_{n-1} z_{n-1}
\end{equation}
where $k=\tau/h$ is the discretized delay. The $z_n$'s are independent Gaussian random numbers with variance 1 which may be generated by a Box-Müller transformation~\cite{BM}. Since a Gaussian distribution is unbound it can happen that the random number~$z_{n-1}$ takes on a very large negative value so that $x_{n+1}$ becomes negative. This overshooting to negative values is a numerical artifact caused by the discretization since in the contiuum limit $x(t)$ is always positive as long as $b\geq 0$. There are several ways to deal with this problem:
\begin{itemize}
\item[(a)] The time increment $h$ may be decreased until the probability for overshooting to negative values is practically zero. However, this method is often inefficient and unsafe.
\item[(b)] Another possiblity is to discard all those updates that would produce negative values. This method is safe but it introduces systematic numerical errors in form of an effective bias towards larger amplitudes of $x_n$.
\item[(c)] The most accurate method is a semi-analytical approach suggested by Dornic \textit{et al.} in Ref.~\cite{Dornic}.
\end{itemize}
%
%%%%%%%%%%%%%%%%%%%%%%%%%%%%%%%%%%%%%%%%%%%%%
\begin{figure}
\centering\includegraphics[width=100mm]{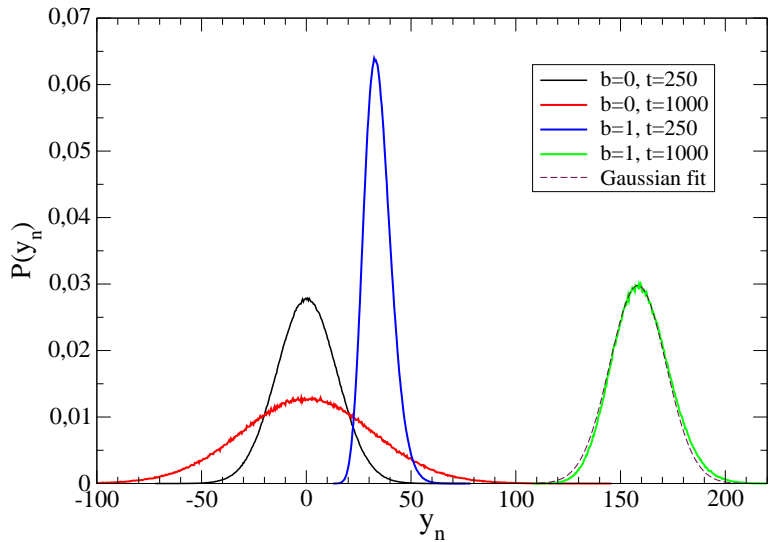}
\caption{\small Distribution of $y_n=\log(x_n)$ of the time series $x_n$ with delay $k=100$ after $n=10^5$ updates for $c=1$ and $a=1/2$. For $b=0$ one obtains a Gaussian distribution centered at the origin. For $b=1$, however, the center of the distribution moves to the right as time proceeds. Moreover, the distribution is no longer Gaussian, instead it becomes skew.}
\label{fig:profiles}
\end{figure}
%%%%%%%%%%%%%%%%%%%%%%%%%%%%%%%%%%%%%%%%%%%%%
%
For the purpose of the present work, where we are primarily interested in qualitative results, the second method with a suitable small time increment~$h$ is sufficiently accurate. 

The iteration starts with the initial condition $x_0=x_1=\ldots=x_{k-1}=1$. Using the step size $h=10^{-2}$ and iterating up to $n=10^5$ steps we monitor the distribution of $x_n$. As observed in Ref.~\cite{Brett}, the $x_n$ are positive and fluctuate strongly over several orders of magnitude. Since the fluctuations are caused by the multiplicative character of the evolution equation, it is reasonable to investigate the distribution of the logarithm
\begin{equation}
y_n = \log(x_n).
\end{equation}
The distributions for $c=1$ and $a=1/2$ are shown in Fig.~\ref{fig:profiles}. For $b=0$ the values of~$y_n$ in an ensemble of independent runs are distributed according to a Gaussian. Because of the special choice $a=1/2$ in the It\^o scheme the distribution is centered at the origin. As expected, the distribution becomes broader as time proceeds. For $b>0$, however, the situation is very different. On the one hand the distribution has a smaller width and moves to the right as time proceeds. On the other hand it is no longer Gaussian, instead it becomes skew (see Fig.~\ref{fig:profiles}). As we will see below, this skewness is one of the hallmarks of dynamical scaling in the KPZ universality class.

%============================================================================================
\section{Relation to the KPZ equation}
%============================================================================================

Let us now demonstrate how the stochastic differential equation~~(\ref{eq:sdeq}) is related to the KPZ equation. We start out with the discretized equation (\ref{eq:original}) with parameters $a$, $b$ and~$h$:
\begin{equation}
\label{eq:origeneral}
x_{n} \;=\; (1+ah)x_{n-1} + bh\,x_{n-k}+\sqrt{h}\, x_{n-1} z_{n-1}
\end{equation}
Consider the mapping $\mathbb N \to \mathbb N \otimes \{0,\ldots k-1\} :\quad n \to (i,j)$
\begin{eqnarray}
i  &=& n \mbox{ div } k \,,\\
j  &=& n \mbox{ mod } k\,,\nonumber
\end{eqnarray}
where `div' and `mod' denote integer division and modulus. As explained in the Introduction, this mapping slices the time series into equidistant blocks of size $k$. The blocks are indexed by $i$ while the index $j$ encodes the position within each block. This transformation maps the one-dimensional time series with non-local (time-delayed) updates onto a 1+1-dimensional model with local (nearest-neighbor) updates of the form~\cite{DHK}
\begin{equation}
\label{eq:twodim}
x_{\scriptscriptstyle i,j} \;=\;  (1+ah)x_{\scriptscriptstyle i,j-1} \,+\,bh\,x_{\scriptscriptstyle i-1,j}+\sqrt{h}\,x_{\scriptscriptstyle i,j-1} z_{\scriptscriptstyle i,j-1}\,,
\end{equation}
where the indices $i$ and $j$ may be interpreted as `time' and `space'. Note that the boundary conditions differ from periodic ones in so far as they involve a shift in the index $i$:
\begin{equation}
x_{\scriptscriptstyle i,-1} \equiv x_{\scriptscriptstyle i-1,k-1}\,\qquad x_{\scriptscriptstyle i,k}\equiv x_{\scriptscriptstyle i+1,0}
\end{equation}
Carrying out the continuum limit and performing a suitable Galilei transformation (see Appendix) the bulk equation~(\ref{eq:twodim}) can be written in terms of differential operators as
\begin{eqnarray}
\label{MNequation}
\nabla_s x(s,r)  &=& (1+\frac{a}{b}) x(s,r) - \frac{1}{2ab}\nabla_r^2 x(s,r) \\
&&+ \frac{1}{b}\Bigl[
1 + \frac{1}{a}\nabla_r + \frac{1}{2a^2}\nabla^2_r 
\Bigr](x(s,r)  \xi(s,r))\,,\nonumber
\end{eqnarray}
where $s$ and $r$ are temporal and spatial coordinates respectively. Apart from the higher-order derivatives in the last term this equation has the same structure as the Langevin equation for 1+1-dimensional systems with multiplicative noise~\cite{Grinstein,Genovese} which is known to exhibit dynamical scaling. This means that the equation is invariant under the scaling transformation
\begin{equation}
r \to \Lambda r\,, \quad s  \to \Lambda^z s
\end{equation}
with a certain dynamical exponent $z>1$. Simple power counting of the leading contributions reveals that $z=2$ above the upper critical dimension $d_c=2$. Moreover, the higher-order derivatives in the last term of Eq.~(\ref{MNequation}) turn out to be irrelevant in the scaling limit and may be discarded. It has widely believed that this applies also to the 1+1-dimensional case. The remaining equation reads
\begin{equation}
\label{MNequation2}
\nabla_s x(s,r)  \;=\; (1+\frac{a}{b}) x(s,r) - \frac{1}{2ab}\nabla_r^2 x(s,r) 
+ \frac{1}{b} x(s,r) \xi(s,r)\,.
\end{equation}
With a Hopf-Cole transformation $x(s,r)=\exp(y(s,r))$ this equation turns into the stochastic differential equation
\begin{equation}
\nabla_s y(s,r) \;=\; 1 + \left(\frac{a}{b}-\frac{1}{2b}\right) - \frac{1}{2ab} \Bigl[ \nabla_r^2 y(s,r) + (\nabla_r y(s,r))^2 \bigr] + \frac{1}{b}\xi(s,r)
\end{equation}
with a non-multiplicative noise. Here the additional term $-\frac{1}{2b}$ is due to the It\^o scheme under the transformation. This equation is known as the Kardar-Parisi-Zhang (KPZ) equation for non-linear surface growth~\cite{KPZ}, where $y(s,r)$ denotes the height of a growing interface at position $r$ at time $s$. Since $s\\simeq t/\tau$ the mapping to the KPZ equation allows us to make the following predictions:
\begin{itemize}
\item The width of a roughening KPZ interface is known to grow as $s^{1/3}$, hence the variance of $y(t)=\log x(t)$ in the original problem should grow as $\sigma^2(t) \sim t^{2/3}$.
\item In the 1+1-dimensional formulation finite-size effects are expected to occur when the spatial correlation length $\xi_\perp \sim s^{1/z}$ becomes comparable with the width of the strip $\tau$, where $z=3/2$ is the dynamical exponent of the KPZ universality class in 1+1 dimensions. Therefore, finite-size effects are expected to set in when $t \sim \tau^{5/2}$.
\item The (rescaled) distribution of $y(t)$ in the range $\tau \ll t \ll \tau^{5/2}$ is no longer given by a Gaussian, instead the distribution becomes skew.
\end{itemize}
The critical exponents as well as the rescaled distribution are universal quantities, i.e., they are fully determined by the underlying field theory of the KPZ class. In a recent breakthrough, Pr\"ahofer and Spohn were able to compute the rescaled height distribution analytically~\cite{prae1,prae2,prae3,prae4}. 

%============================================================================================
\section{Numerical validation}
%============================================================================================

%%%%%%%%%%%%%%%%%%%%%%%%%%%%%%%%%%%%%%%%%%%%%
\begin{figure}
\centering\includegraphics[width=130mm]{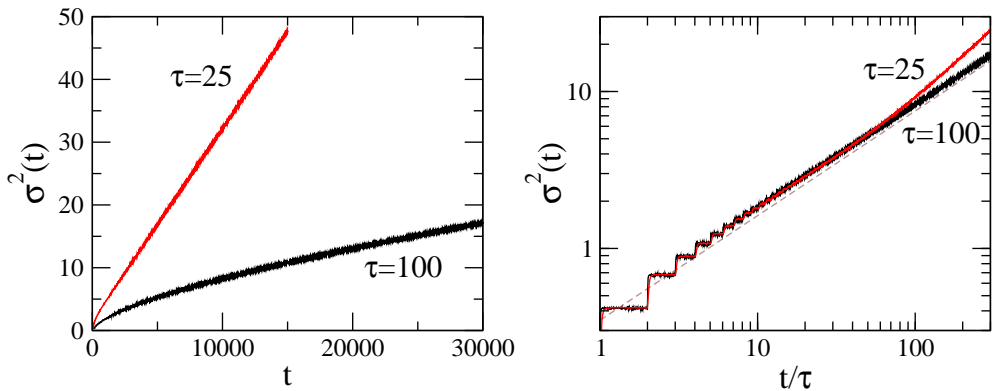}
\caption{\small Variance of $y(t)=\log(x(t))$ in an ensemble of independent iterations for $b=c=1$ and two different values of the delay $\tau$ (see text). The dashed line indicates the slope $2/3$.}
\label{fig:width}
\end{figure}
%%%%%%%%%%%%%%%%%%%%%%%%%%%%%%%%%%%%%%%%%%%%%

In order to verify these predictions, we iterated the discretized version of the original stochastic differential equation numerically. Throughout all simulations we used the choice $b=c=1$. In order to avoid floating-point overflows the parameter $a$ was adjusted at the beginning of each simulation by an iterative procedure such that $y(t) = \log x(t)$ does not grow on average. We produced three data sets for step sizes $h=0.02$, $0.01$, and $0.005$. On the time scales of interest these data sets coincide within statistical errors, meaning that deviations due to the discretization can be neglected. 

The left panel of Fig.~\ref{fig:width} shows the variance $\sigma^2(t) = \langle (y(t)-\langle y(t) \rangle)^2\rangle$ as a function of $t$ for two different delays $\tau=25$ and $\tau=100$. As expected, the variance grows faster for small $\tau$. The right panel shows the same data in a double-logarithmic representation as a function of the quotient $s\simeq t/\tau$. For small $t/\tau$ the two curves collapse and one can clearly see the block structure in form of a staircase. The curve for the large delay $\tau=100$ displays a clean power law with the slope $0.665(5)$, confirming the KPZ roughening-exponent $2\beta = 2\alpha/z = 2/3$ in 1+1 dimensions. The curve for the smaller delay $\tau=25$ begins to deviate from this power law at $t/\tau \approx 100$. These deviations are caused by finite-size effects and are expected to set in at a time scale $t \sim \tau^{5/2}$.

%%%%%%%%%%%%%%%%%%%%%%%%%%%%%%%%%%%%%%%%%%%%%
\begin{figure}
\centering\includegraphics[width=130mm]{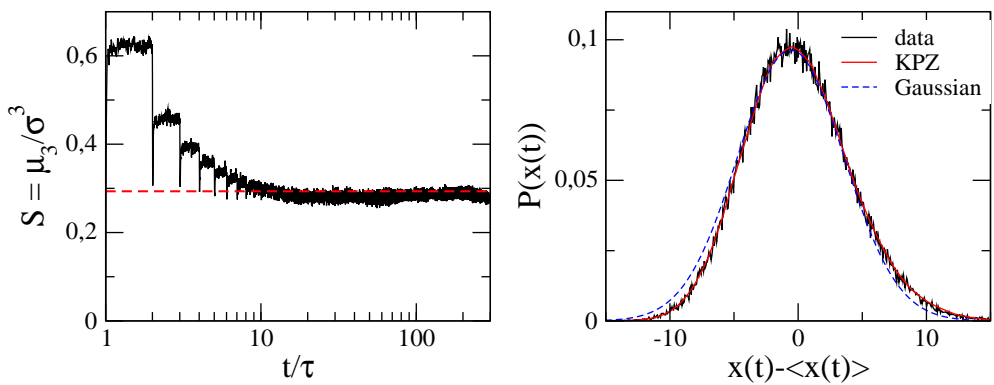}
\caption{\small Left: Skewness of the probability distribution of $y(t)$ as a function of time for $b=c=1$ and $\tau=100$. The theoretical prediction for KPZ roughening is marked as a dashed line. Right: Form of the distribution at time $t=30000$ for the same parameters compared to a high-precision profile of a KPZ growth process. To demonstrate the skewness we fitted a Gaussian (dashed line) in the center.}
\label{fig:skewness}
\end{figure}
%%%%%%%%%%%%%%%%%%%%%%%%%%%%%%%%%%%%%%%%%%%%%

In order to substantiate these findings even further, we monitored the skewness
\begin{equation}
S(t) \;=\; \frac{\mu_3(t)}{\sigma(t)^3} \;=\; \frac{\langle ( y(t)-\langle y(t) \rangle)^3 \rangle }
{\langle ( y(t)-\langle y(t) \rangle)^2 \rangle^{3/2}}
\end{equation}
as a function of time, using the parameters $b=c=1$ and $\tau=100$. A Gaussian distribution would yield the value $S=0$ while a KPZ profile is characterized by a non-zero value. In more recent simulations~\cite{krug} this value was estimated by $S_{\rm \scriptscriptstyle KPZ}=0.28(4)$ while Pr{\"a}hofer and Spohn find the analytical result~\cite{prae1,prae2,prae3,prae4}
\begin{equation}
S_{\rm \scriptscriptstyle KPZ}\approx 0.2935.
\end{equation}
As shown in the left panel of Fig.~\ref{fig:skewness}, the skewness of the distribution quickly saturates at a constant which is compatible with the theoretical prediction (dashed line). The right panel shows the actual form of the distribution at time $t=30000$ in comparison with a (rescaled) high-precision profile from a simulation of the so-called single-step model~\cite{ginelli} which is known to belong to the KPZ universality class. As can be seen, both curves coincide perfectly and deviate significantly from a Gaussian (shown as a dashed line).

To summarize, there is strong numerical evidence that $y(t)=\log(x(t))$ is indeed distributed in the same way as the heights of a roughening KPZ interface, supporting the analytical derivation in Sect. 3.

%============================================================================================
\section{Conclusions}
%============================================================================================

In this paper we have shown how a one-dimensional system with time-delayed dynamics governed by Eq.~(\ref{eq:sdeq}) can
be mapped onto a 1+1 dimensional stochastic model with local dynamics in the KPZ universality class. Our analytical results
are validated by numerical simulations. The example discussed here involves two delays. One of them is local and arises from
the discretization of the time derivative, while the other one is nonlocal. One interesting question is whether the scheme here
presented can be extended to more general systems where both delays are nonlocal
\begin{equation}
x(t) = a \,x(t-\tau_1) + b \, x(t-\tau_2 ) + c\,x(t)\,\xi(t)\,.
\end{equation}
Some recent results show that for such stochastic dynamical rules it is always possible to find an equivalent
1+1 dimensional system with local interactions but time-delayed boundary conditions, the only condition
being that the discretized counterparts of $\tau_1$ and $\tau_2$ be
coprime (two numbers are coprime if they have no common divisor other than 1). Hence we
conjecture that the KPZ universality class of growing interfaces should encompass a much broader class
of systems whose dynamics is time-delayed. Work in this direction is currently under progress~\cite{DHK}.
%============================================================================================
\section*{Acknowledgements}
The authors would like to thank C. Brettschneider and W. Kinzel for helpful discussions.
S.R.D. would like to thank the Alexander von Humboldt Foundation for the financial support and the
Computational Physics Group at W\"urzburg for the hospitality.
%============================================================================================
\appendix
\section{Continuum limit}
%============================================================================================

In this appendix we show how the continuum limit is carried out. Starting point is the 1+1-dimensional discretized equation 
\begin{equation}
\label{eq:twodimapp}
x_{\scriptscriptstyle i,j} \;=\;  (1+ah)x_{\scriptscriptstyle i,j-1}\,+\,bh\,x_{\scriptscriptstyle i-1,j}+\sqrt{h}\, x_{\scriptscriptstyle i,j-1} z_{\scriptscriptstyle i,j-1}\,.
\end{equation}
To perform the continuum limit we introduce continuous coordinates $s,r$ by
\begin{equation}
s = \hs\ i\,, \quad r=\hr\ j\,
\end{equation}
with step sizes $\hs$ and $\hr$ to be determined later. The discretized field and the noise are approximated by smooth functions 
\begin{equation}
x_{i,j}\longrightarrow x(s,r)\,;\qquad z_{\scriptscriptstyle i,j}\longrightarrow
\sqrt{\hs\hr}\,\xi(s,r)\,,
\end{equation}
where $\xi(r,s)$ is a Gaussian white noise with the correlations 
\begin{equation}
\langle \xi(r,s)\xi(r',s')\rangle  \;=\;\delta(r-r') \delta(s-s')\,.
\end{equation}
With this notation the equation above takes the form
\begin{eqnarray}
x(s,r) &=&  (1+ah) \, x(s,r-\hr) + bh\, x(s-\hs,r) \\
&&+ \sqrt{h\hr\hs}\,x(s,r-\hr)\xi(s,r-\hr)\,,\nonumber
\end{eqnarray}
This equation may be written equivalently as
\begin{eqnarray}
x(s,r) &=& (1+ah) e^{-\hr\nabla_r} x(s,r) +bh\,e^{-\hs\nabla_s} x(s,r) \\
&&+ \sqrt{h\hr\hs}\,e^{-\hr\nabla_r} [x(s,r) \xi(s,r)]\,.\nonumber
\end{eqnarray}
In order to compensate for a possible drift we apply a Galilei transformation
\begin{equation}
r \to r + v s \,, \qquad \nabla_s \to \nabla_s - v \nabla_r\,.
\end{equation}
With this transformation the equation becomes
\begin{eqnarray}
e^{-\hs v\nabla_r} x(s,r) &=& (1+ah) e^{-(\hr+v\hs)\nabla_r} x(s,r) +bh\,e^{-\hs\nabla_s} x(s,r)
\\ 
&&+ \sqrt{h\hr\hs}\,e^{-(\hr + v\hs)\nabla_r} [x(s,r) \xi(s,r)]\,.
\end{eqnarray}
Expanding to first order in $s$ and to second order in $r$ we obtain
\begin{eqnarray}
&&bh\hs\nabla_s x(s,r) \;=\\
&&\quad h(a+b)x(s,r) - \Bigl[(1+ah)\hr + ah\hs v\Bigr] \nabla_r x(s,r) \nonumber\\
&&\quad+ \frac12 \Bigl[ (1+ah)(\hr^2+2\hr\hs v)+ah\hs^2v^2 \Bigr] \nabla_r^2  x(s,r) \nonumber\\
&&\quad + \sqrt{h\hr\hs}\Bigl(1-(\hr+v\hs)\nabla_r + \frac12 (\hr +v\hs)^2 \nabla_r ^2\Bigr) [x(s,r)  \xi(s,r)]\nonumber\,.\nonumber
\end{eqnarray}
Now we adjust the velocity $v$ in such a way that the drift (the second term on the rhs) vanishes. This can be done by setting
\begin{equation}
v = -\frac{\hr}{\hs}\Bigl(\frac{1+ah}{ah}\Bigr)\,,
\end{equation}
leading to the equation
\begin{eqnarray}
bh\hs \nabla_s x(s,r) &=& h(a+b) x(s,r) - \frac{(1+ah)\hr^2}{2ah} \nabla_r^2  x(s,r)\\
&& +  \sqrt{h\hr\hs} \Bigl[
1 + \frac{\hr}{a h}\nabla_r + \frac12\left(\frac{\hr}{a h}\right)^2\nabla^2_r 
\Bigr]  (x(sr)\xi(s,r))\nonumber
\end{eqnarray}
In the limit $\hr=h\to 0$ and taking $\hs=1$ one finally obtains
\begin{eqnarray}
\nabla_s x(s,r)  &=& (1+\frac{a}{b}) x(s,r) - \frac{1}{2ab}\nabla_r^2 x(s,r) \\
&&+ \frac{1}{b}\Bigl[
1 + \frac{1}{a}\nabla_r + \frac{1}{2a^2}\nabla^2_r 
\Bigr](x(s,r)  \xi(s,r))\nonumber
\end{eqnarray}
which is understood to be iterated in the It\^o sense. 

%============================================================================================
% References
%============================================================================================

\end{document}